\documentclass{natureprintstyle}
\usepackage{graphicx,amsmath}
\usepackage{astjnlabbrev-nature} 

\usepackage{amssymb}
\title{The exclusion of a significant range of ages in a massive star cluster}

\author{Chengyuan Li$^{1,2,3}$, Richard de Grijs$^{1,2}$ and Licai Deng$^{3}$}
\begin{document}

\maketitle

\begin{affiliations}
 \item Kavli Institute for Astronomy \& Astrophysics, Peking
   University, Yi He Yuan Lu 5, Hai Dian District, Beijing 100871,
   China\\
 \item Department of Astronomy, Peking University, Yi He Yuan Lu 5,
   Hai Dian District, Beijing 100871, China\\
 \item Key Laboratory for Optical Astronomy, National Astronomical
   Observatories, Chinese Academy of Sciences, 20A Datun Road,
   Chaoyang District, Beijing 100012, China\\
\end{affiliations}

\begin{abstract} 
Stars spend most of their lifetimes on the main sequence in the
Hertzsprung--Russell diagram. The extended main-sequence turn-off
regions -- containing stars leaving the main sequence after having spent 
all of the hydrogen in their cores -- found in massive (more than a few 
tens of thousands of solar masses),
intermediate-age (about one to three billion years old) star
clusters\cite{Mack07,Mack08,Milo09,Rube10,Goud11,Kell12,Rube13,Li14}
are usually interpreted as evidence of cluster-internal age spreads
of more than 300 million years\cite{Mack08,Rube10,Goud11}, although young
clusters are thought to quickly lose any remaining star-forming fuel
following a period of rapid gas expulsion on timescales of order
$10^7$ years\cite{Bast06,Longmore14}. Here we report that the stars
beyond the main sequence in the two billion-year-old cluster NGC 1651,
characterized by a mass of $\sim 1.7 \times 10^5$ solar
masses\cite{Milo09}, can be explained only by a single-age stellar
population, even though the cluster has clearly extended main-sequence
turn-off region. The most plausible explanation for the extended
main-sequence turn-offs invokes the presence of a population of
rapidly rotating stars, although the secondary effects of the
prolonged stellar lifetimes associated with such a stellar-population
mixture are as yet poorly understood. From preliminary analysis of previously 
obtained data, we find that
similar morphologies are apparent in the Hertzsprung--Russell diagrams
of at least five additional intermediate-age star
clusters\cite{Mack08,Milo09,Goud11,Piat14}, suggesting that an
extended main-sequence turn-off does not necessarily imply the
presence of a significant cinternal age dispersion.
\end{abstract}

We obtained archival {\sl Hubble Space Telescope}/Wide Field Camera-3
observations of the NGC 1651 field in the F475W (``$B$'') and F814W
(``$I$'') broad-band filters (Methods). The corresponding
colour--magnitude diagram, that is, the observational counterpart of the
Hertzsprung--Russell diagram, is shown in Fig. 1. When stars have
exhausted their core hydrogen supply, hydrogen fusion continues in a
shell outside the stellar core. At this stage, stars leave the main
sequence and evolve onto the subgiant branch. The colour--magnitude
diagram of NGC 1651 exhibits a clearly extended main-sequence turn-off
and a very narrow subgiant branch. This is surprising, given the
corresponding, far-reaching implications for our interpretation of
such extended turn-offs in the context of star cluster evolution.

Star clusters more massive than a few tens thousands of solar masses were,
until recently, considered single-generation (``simple'') stellar
populations. It was thought that all of their member stars had formed
approximately simultaneously from molecular gas originally confined to
a small volume of space. As a consequence, all cluster stars would
thus have similar ages, a very narrow range in chemical composition
and individual stellar masses that followed the initial mass function, that is, 
the stellar mass distribution at the time of star birth. 
In the past decade, however, consensus has emerged that
massive star clusters are no longer ideal simple stellar
populations\cite{Piot05,Piot07,Soll07,Milo08,Lee09,Milo12,Piot13}.
Deviations from the simple-stellar-population model in resolved star
clusters are most readily discerned by reference to their
colour--magnitude diagrams, and in particular to their main-sequence
turn-off regions.

\begin{figure*}
\hspace{-0.4cm}\centering
\includegraphics[width=1.6\columnwidth]{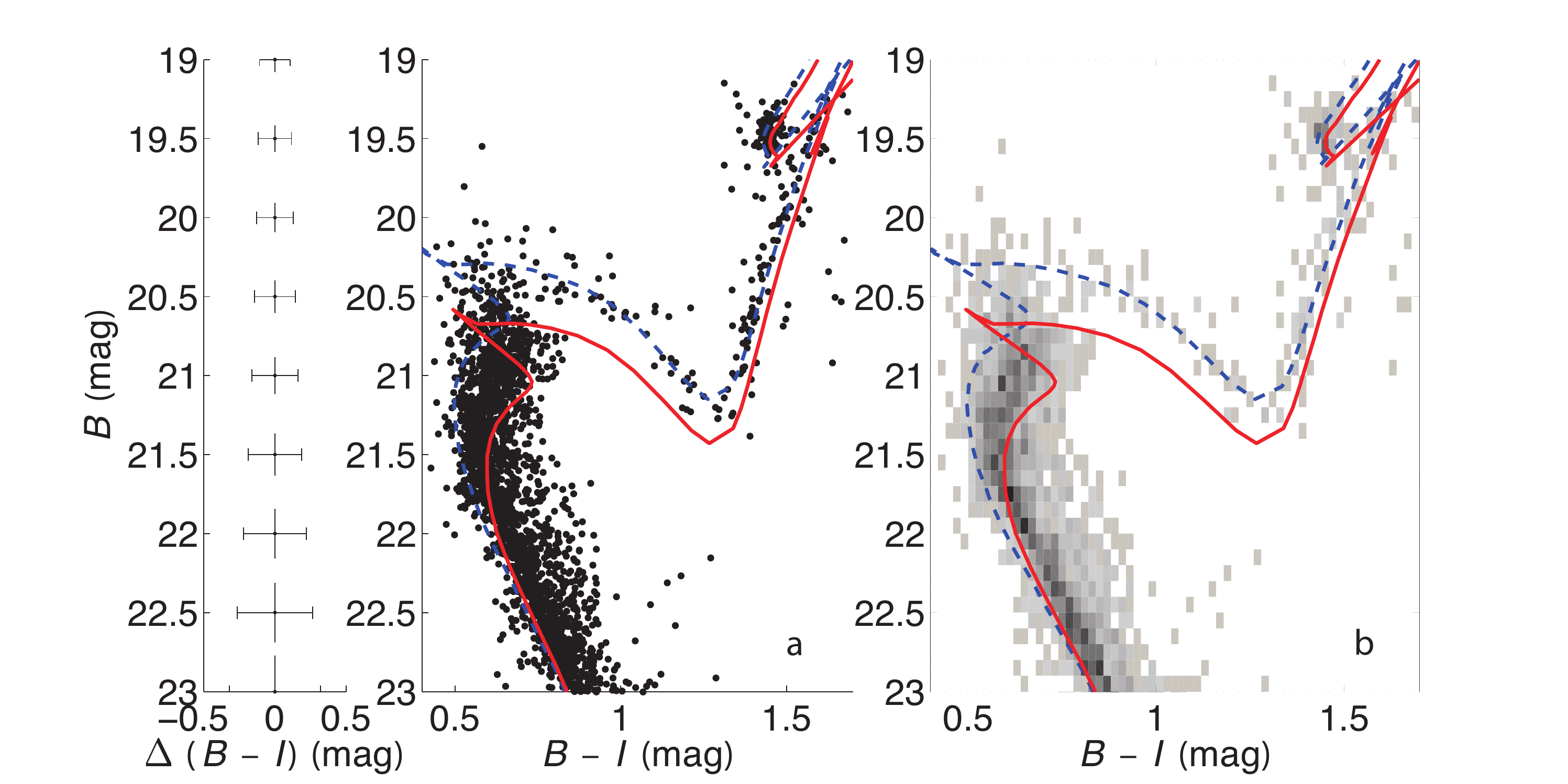}
\begin{center}
\caption{{\bf NGC 1651's stellar distribution in colour--magnitude
  space.} a, Colour--magnitude diagram, including typical $3\sigma$
  photometric uncertainties. The blue dashed and red solid lines
  represent isochrones for $\log (t \mbox{ yr}^{-1}) = 9.24$ and 9.34,
  respectively. b, Corresponding number-density (``Hess'') diagram.}
\end{center}
\label{F1}
\end{figure*} 

Taking a simplistic, direct approach, we obtain best fits to the blue and red edges of
the extended turn-off by matching the best set of
theoretical stellar isochrones\cite{Mari08} available at present to the observed stellar
distribution. The best-fitting isochrones bracketing the data range
from $\log[t \,({\rm yr})] = 9.24$ to $\log[t \,({\rm yr})] = 9.34$ (where $t$ represents
the stellar population's age), for a stellar metal (iron) abundance,
[Fe/H] $= -0.52$ dex (ref. 20),a reddening $E(B - V) = 0.11$ mag
and a distance modulus, $(m - M)_0 = 18.46$ mag (ref.21).Figure 1
shows the `cleaned' colour--magnitude diagram (Methods). The
lines represent the best-fitting theoretical isochrones covering the
cluster's extended main-sequence turn-off. Although the latter is
well-described by adoption of an age dispersion of approximately 450
Myr, the cluster's subgiant-branch stars are predominantly confined to
the youngest isochrone. The 15 subgiant-branch stars in the NGC 1651
core region (with a radius of $\leq20$ arcsec $\equiv 5$ pc; see Methods) are confined
to an even narrower distribution along the subgiant branch than is the
full sample of 38 stars selected using the box in Fig. 2a, c. This
indicates that the narrow width of the subgiant branch does not depend
on position in the cluster. However, a 450 Myr age spread would {\it also}
require a significant broadening of the cluster's subgiant
branch. This is why our discovery of a subgiant branch in
NGC 1651 with a very narrow stellar distribution is surprising, which thus 
immediately leaves us with a troublesome conundrum.

\begin{figure}[h!]
\centering
\includegraphics[width=0.95\columnwidth]{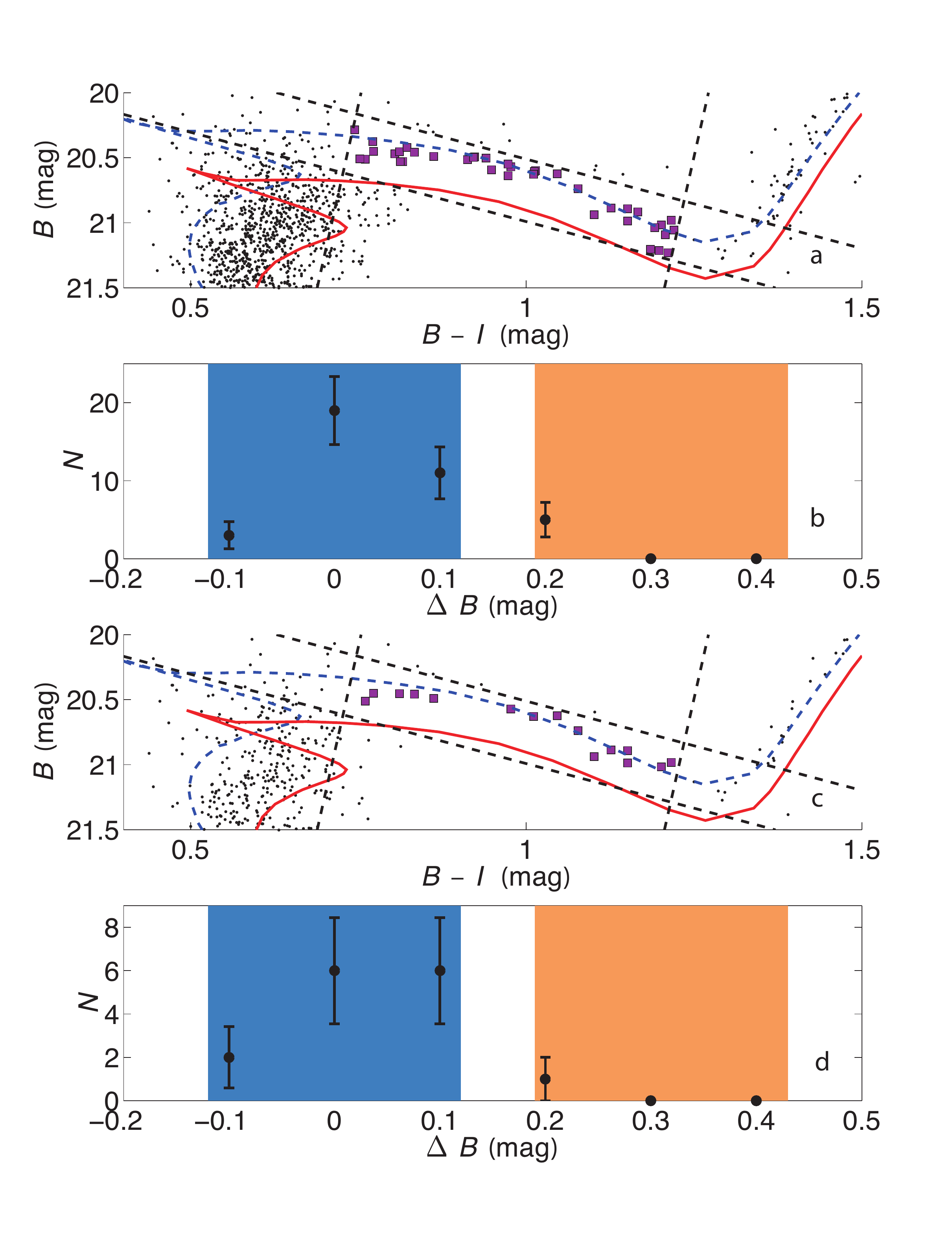}  
\begin{center}
\caption{{\bf Comparison of the observed stellar distribution with the expectations of a 450 Myr spread in cluster internal age.} a, Region of the colourÐmagnitude diagram covering the extended turn-off and the subgiant branch (indicated by the black dashed lines; purple squares, subgiant-branch stars). The blue dashed and red solid isochrones are as in Fig. 1. b, Number distribution, N (including 1$\sigma$ standard deviations), of the deviations in
magnitude, $\Delta$B, of our subgiant-branch sample from the youngest and oldest isochrones (light blue and orange backgrounds, respectively). c, d, As in
a (c) and b (d), but for subgiant-branch stars in the cluster core, that is, for stars located at radii of $\leq$20 arcsec.}
\end{center}
\label{F2}
\end{figure} 

To assess the association of our subgiant-branch stars with either the
youngest or oldest isochrones, we first adopt the $\log[t \,({\rm yr})] = 9.24$ 
isochrone as our baseline and calculate the
individual deviations, $\Delta B$ (mag), for all
subgiant-branch stars. We subsequently adopt the $\log[t \,({\rm yr})] = 9.34$ 
isochrone as our fiducial locus. The blue and
orange regions in Fig. 2b, d correspond to the typical deviations
expected for subgiant-branch stars associated with the young and old
isochrones, respectively, assuming a 3$\sigma$ magnitude dispersion of
$\Delta B = 0.12$ mag. Thirty of the 38 stars (14 of 15 stars in the
core) are associated with the young isochrone. Only a single
subgiant-branch star, located outside the cluster's core region, might
statistically be associated with the region in parameter space defined
by the old isochrone. If we directly use the observed spread of these
stars in the colour--magnitude diagram to derive a maximum likely age
spread, $\Delta t$, we conclude that $\Delta t \le 160$ Myr for the full sample and that 
$\Delta t \le 80$ Myr for the core sample (Methods).

\begin{figure}  
\centering 
\includegraphics[width=\columnwidth]{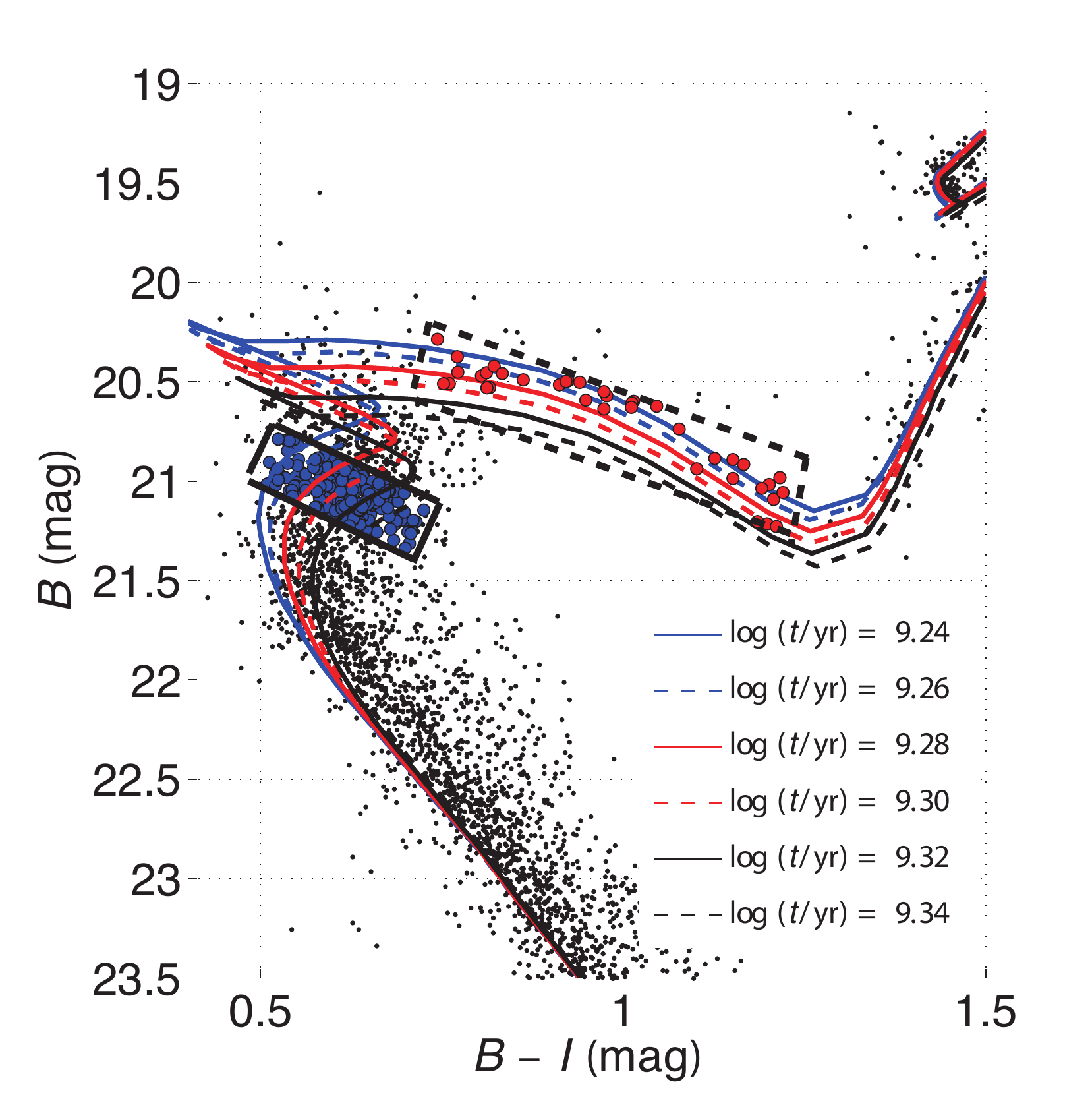}  
\begin{center}
\caption{{\bf Comparison of the numbers of stars in NGC 1651 at selected
  evolutionary stages.} Blue points: `Typical' main-sequence turn-off stars
  used as basis for the comparison; red points: Comparison sample of
  subgiant-branch stars. Isochrones for different ages are also shown
  (see key).}
\end{center}
\label{F3}
\end{figure} 

If the cluster's stellar population were characterized by an age
dispersion, this would naturally produce a broadened subgiant
branch. Using Fig. 3, we quantitatively assess the probability of the
presence of a genuine internal age dispersion. We calculated the
number-density distributions of both ``typical'' extended turn-off
stars (blue bullets in Fig. 3) and the cluster's subgiant-branch
stars, adopting differently aged isochrones (see figure legend). The
resulting distributions are indeed significantly different, as shown
in Fig. 4 (a, b). Whereas the extended turn-off stars clearly exhibit
a spread from $\log[t \,({\rm yr})]  = 9.24$ to 9.34, the
subgiant-branch stars are almost all associated with the young
isochrone. Once again, this invalidates the presence of a genuine age
spread within the cluster.

\begin{figure}  
\centering 
\includegraphics[width=\columnwidth]{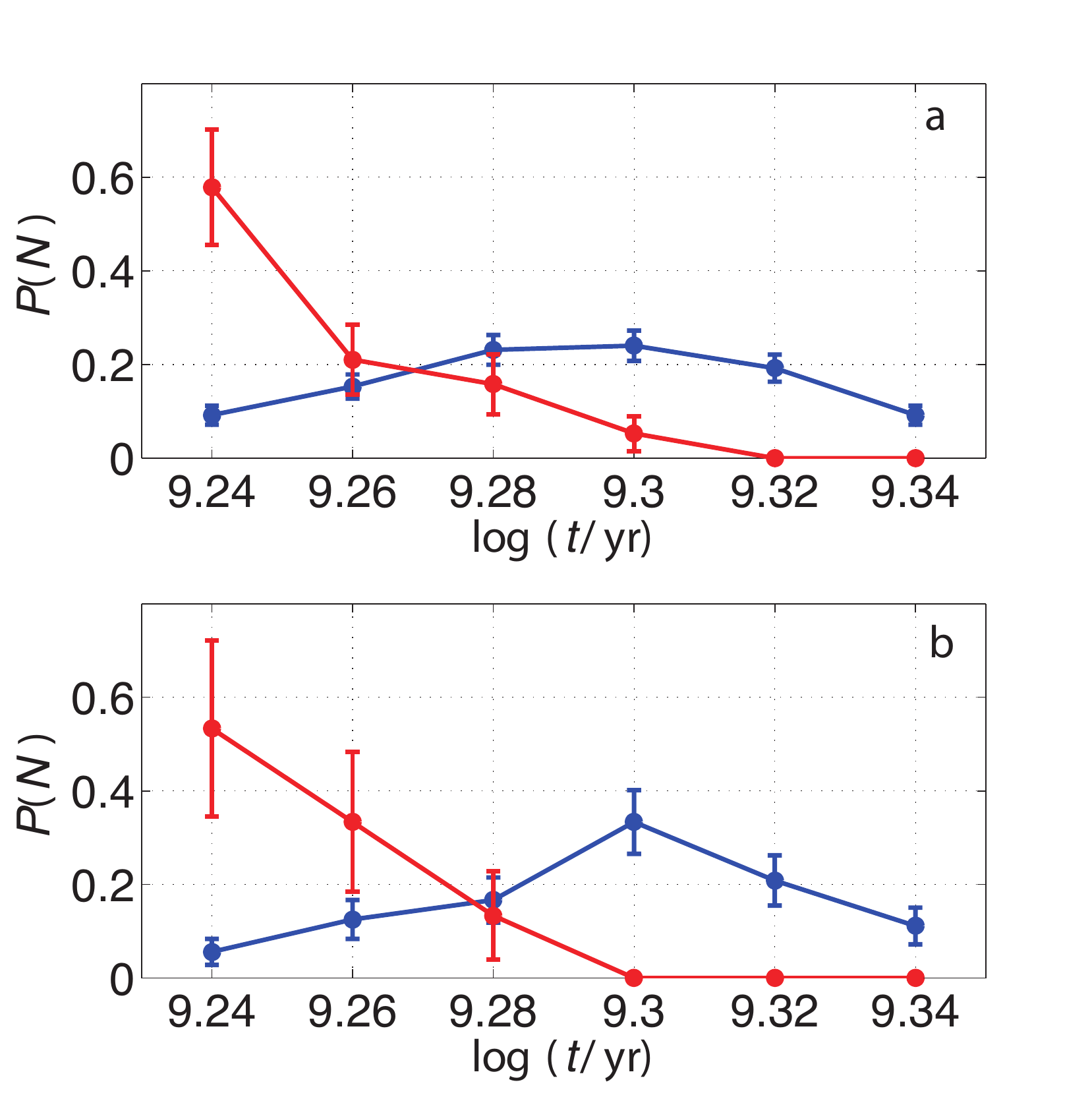}  
\begin{center}
\caption{{\bf Expected age distributions resulting from the cluster's
  turn-off and subgiant-branch stars.} a, Number-density distribution,
  $P(N)$, as a function of age, including their 1$\sigma$ standard
  deviations. Blue: extended main-sequence turn-off stars. Red:
  Subgiant-branch stars. b, As panel a, but for the cluster core
  region.}
\end{center}
\label{F4}
\end{figure} 

It is imperative to probe beyond the extended main-sequence turn-off
to fully understand the evolution of massive clusters at ages in
excess of 1 Gyr. Subgiant-branch stars will not yet have
experienced significant mass loss, which would further complicate our
interpretation of, for example, the morphology of the upper end of the
red-giant branch and of the red clump. Investigation of the
subgiant-branch morphology thus offers clean insights into the extent
to which intermediate-age clusters resemble true simple stellar
populations, unimpeded by effects owing to unresolved binary
systems\cite{Hu10,Li13} or the possible presence of a population of
rapidly rotating stars\cite{Bast09,Li12,Yang13,Li14}, both of which
complicate our interpretation of the nature of the observed extended
main-sequence turn-offs. Unresolved binary systems will broaden the
turn-off towards lower magnitude (brighter stars), but they will not cause a
reddening of this region\cite{Mack08,Milo09}. Our discovery of a very
narrow subgiant branch in NGC 1651 implies that the impact of binary
systems is negligible.

The possible presence of a population of rapidly rotating stars may
also complicate our interpretation of the observed, extended
main-sequence turn-offs in intermediate-age
clusters\cite{Bast09,Li14}. Moreover, because of the conservation of
angular momentum, any fast rotators on the main sequence are (naively)
expected to quickly slow down when they expand and evolve onto the
subgiant branch. However, in practice the contribution to the
subgiant-branch morphology from a population of rapidly rotating stars
is complex, given that fast stellar rotation facilitates prolonged
main-sequence lifetimes\cite{Gira13}. The presence of such stars may,
in fact, also cause a subgiant-branch split\cite{Geor14}, driven by
the resulting extended characteristic stellar mass range and its
corresponding range in evolutionary timescales. However, the
importance of this latter scenario strongly depends on the prevailing
mixing efficiency\cite{Yang13}. For sufficiently small mixing
efficiencies, the turn-off region will be broadened while the subgiant
branch will remain relatively narrow (Methods).

Nevertheless, the observed narrow subgiant-branch width provides the
strongest evidence yet that NGC 1651 cannot have undergone star
formation for any significant, sustained length of time. This thus
implies that an extended main-sequence turn-off in the
colour--magnitude diagram of an intermediate-age massive cluster does
not necessarily imply the presence of a significant, $\gtrsim 100$ Myr
 age dispersion. NGC 1651 is so far unique, because its
subgiant branch is the narrowest yet discovered and discussed for any
cluster characterized by an extended main-sequence turn-off, thus
supporting its nature as a genuine simple stellar population (for
chemical composition-related arguments, see Methods). In retrospect,
other intermediate-age clusters have been found that exhibit extended
main-sequence turn-offs but which also exhibit very narrow subgiant
branches, including NGC 1783.\cite{Mack08,Milo09,Goud11}, NGC 1806\cite{Mack08,Milo09,Goud11}, NGC 1846\cite{Mack08,Milo09,Goud11}, 
NGC 2155\cite{Piat14} and SL 674\cite{Piat14}. The results highlighted here
have left us with an as yet unresolved puzzle regarding the evolution
of young and intermediate-age massive star clusters. This is
troublesome, since star clusters are among the brightest stellar
population components in any galaxy; they are visible to much greater
distances than individual stars, even the brightest. Understanding
star cluster composition in detail is therefore imperative for
understanding the evolution of galaxies as a whole.

\begin{addendum}
 \item[Acknowledgements] We thank Selma de Mink, Yang Huang and
   Xiaodian Chen for discussions and assistance. Partial financial
   support for this work was provided by the National Natural Science
   Foundation of China through grants 11073001,11373010 and 11473037.
 \item[Author Contributions] C.L., R.d.G. and L.D. jointly designed
   and coordinated this study. C.L. performed the data
   reduction. C.L. and R.d.G. collaborated on the detailed
   analysis. L.D. provided ideas that improved the study's
   robustness. All authors read, commented on and jointly approved
   submission of this article.
 \item[Author Information] Correspondence and requests for materials
   should be addressed to C.L. (email: joshuali@pku.edu.cn)
 \item[Competing Interests] The authors declare that they have no
   competing financial interests.
\end{addendum}

\newpage
\section*{\large Methods}

\section*{Observations and Reduction}

The data sets of NGC 1651 were obtained as part of {\it Hubble Space
  Telescope} programme GO-12257 (Principal Investigator: L. Girardi),
using the Wide Field Camera-3 (WFC3). Both clusters were observed
through the F475W and F814W filters (with central wavelengths of 475
nm and 814 nm, respectively), which roughly correspond to the
Johnson--Cousins $B$ and $I$ bands, respectively. Two images with long
exposure times of 1440 s and 1430 s in the $B$ and $I$ bands,
respectively. Two images with short exposure times of
720 s and 700 s, respectively, were obtained. We used the {\sc
  IRAF/DAOPHOT} software package to perform point-spread-function
photometry\cite{Davi94s}.

The photometric catalogues pertaining to the long- and short-
exposure-time images were combined. We carefully cross-referenced both
catalogues to avoid duplication of objects in the combined output
catalogue. For stars in common between both catalogues, we adopted the
generally more accurate photometry from the long-exposure catalogue
for inclusion in the output master catalogue, except in the magnitude
range where the long-exposure image could be affected by saturation
(for example for stars on the upper red-giant branch or blue
stragglers).\cite{Li13}

\section*{Determination of the cluster region}

We divided the stellar spatial distribution into 20 bins along both
the right ascension ($\alpha_{\rm J2000}$) and declination
($\delta_{\rm J2000}$) axes. Using a Gaussian function to fit the
stellar number-density distribution in each direction, we determined
the closest coincidence of both Gaussian peaks as the cluster centre:
$\alpha_{\rm J2000} = 04^{\rm h}37^{\rm m}32.16^{\rm s}
(69.3843^{\circ}), \delta_{\rm J2000} = -70^{\circ} 35'08.88''
(-70.5858^{\circ})$. The centre position compares very well with
previous determinations. For instance, NASA's Extragalactic Database
(http://ned.ipac.caltech.edu) lists $\alpha_{\rm J2000} = 04^{\rm
  h}37^{\rm m}32.3^{\rm s}, \delta_{\rm J2000} = -70^{\circ} 35'09''$,
while SIMBAD (http://simbad.u-strasbg.fr/) gives $\alpha_{\rm J2000} =
04^{\rm h}37^{\rm m}31.1^{\rm s}, \delta_{\rm J2000} = -70^{\circ}
35'02''$, compared with the NGC/IC Project's
(http://www.ngcicproject.org/realskyview/N1600-N1699.txt) $\alpha_{\rm
  J2000} = 4.625750^{\rm h} \equiv 69.38625^{\circ}, \delta_{\rm
  J2000} = -70.585560^{\circ}$.

The complete data set for this cluster is composed of a combination of
two WFC3 images. We used a Monte Carlo-based method to estimate the
areas of rings of different clustercentric radii (all radii were measured 
from the centre of the cluster). Specifically, we
calculated the total area of the region covered and subsequently
generated millions of points that were homogeneously distributed
across the full region. We then calculated the number fraction of
points located in each ring to the total number of points. We used
this fraction, multiplied by the total area to represent the specific
area of each ring. The number of stars in each ring is $N(R)/A(R)$,
where $N(R)$ is the number of observed stars located in a ring with
radius $R$ and $A(R)$ is the corresponding area of the ring.

We next calculated the total brightness of stars in each ring, $f(R) =
\sum_{N}10^{(B_{N} - (m-M)_0)/(-2.5)}$, where $N$ is the number of
stars located in the ring of interest and $(m-M)_0 = 18.46$ mag is the
adopted distance modulus. The brightness density is then $\rho_{f}(R)
= f(R)/A(R)$, which corresponds to a surface brightness, $\mu(R) =
-2.5 \log [\rho_{f}(R)] + 18.46$. Because NGC 1651 is an
intermediate-age star cluster, we represent its brightness profile
by\cite{Elso89s,Mack03s}
\begin{equation}
  f(r) = f_0\left(1+\frac{r^2}{a^2}\right)^{-\gamma/2},
\end{equation}
where $f_0$ is the central surface brightness. The measure of the core
radius, $a$, and the power-law index $\gamma$ are linked to the King
core radius, $r_{\rm c}$, through
\begin{equation}
  r_{\rm c} = a(2^{2/\gamma}-1)^{1/2}.
\end{equation}

The cluster's radial profile, including the 1$\sigma$ photometric
uncertainties due to Poisson noise, as well as the best-fitting
theoretical profile, are shown in Extended Data Fig. 1.

\setcounter{figure}{0}

\begin{figure*}
  \centering
  \includegraphics[width=1.8\columnwidth]{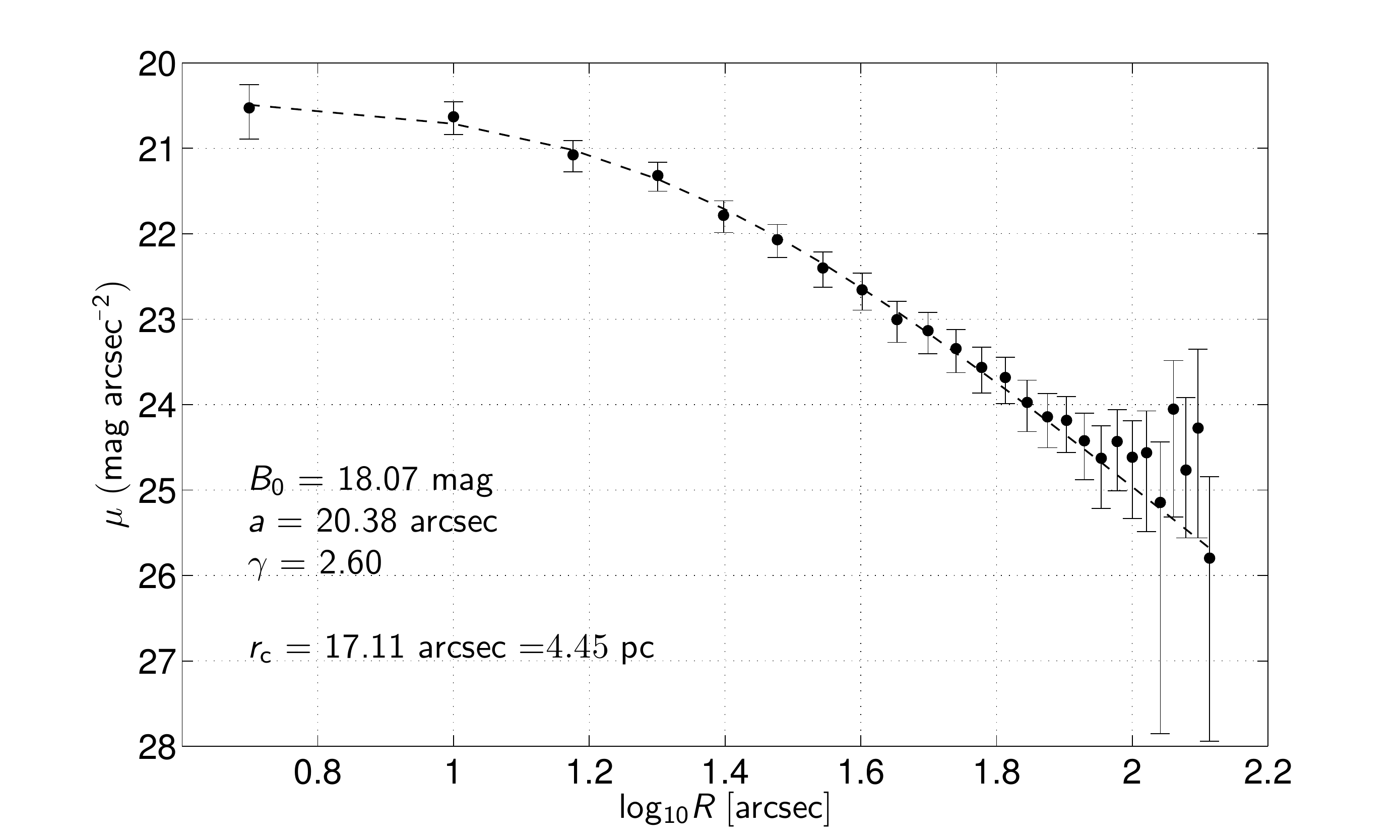}
\caption{{\bf Radial brightness-density profile of NGC 1651.} The 1$\sigma$
  uncertainties shown are owing to Poisson noise.}
  \label{S0}  
\end{figure*}

\section*{Field-star decontamination}

The {\sl Hubble Space Telescope}/WFC3 images cover a very large
region, allowing us to investigate the entire cluster, as well as a
neighbouring field region. Based on the radial density profile in
Extended Data Fig. 2, we determined that for $R \ge 85$ arcsec, the
cluster brightness becomes indistinguishable from the background
noise. We hence selected the region characterized by $R \ge 85$ arcsec
as our comparison field region for the purposes of field-star
decontamination. Taking into account the standard deviation of the
field-star magnitudes, we concluded that the most representative
cluster region has a radius of $R = 75$ arcsec. We statistically
field-star decontaminated this cluster region. Using a Monte Carlo
approach, we estimated that the comparison field region covers 46.7\%
of the cluster region.

\begin{figure*}
  \centering
  \includegraphics[width=1.8\columnwidth]{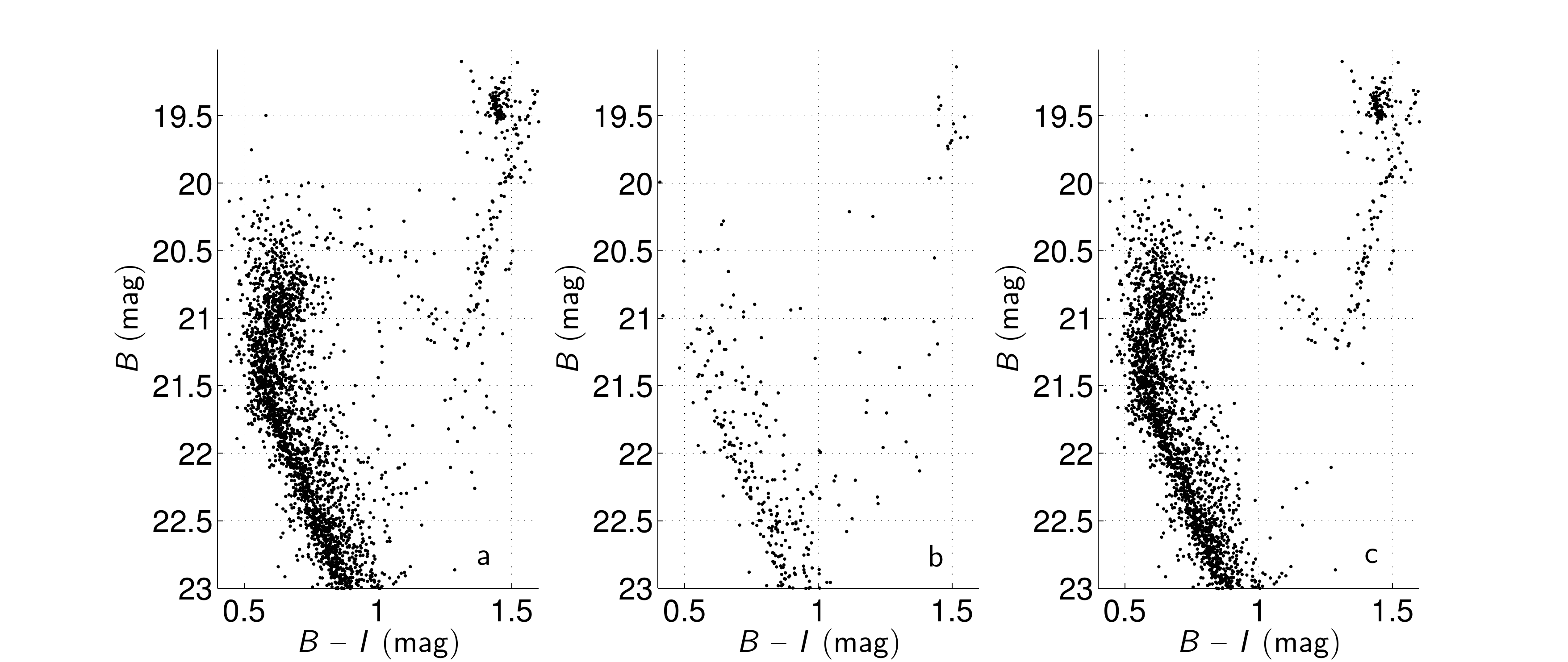}
\caption{{\bf Background decontamination.} a, Original colour--magnitude
  diagram of NGC 1651. b, Field-star colour--magnitude diagram. c,
  Field-star decontaminated NGC 1651 colour--magnitude diagram.}
  \label{S1}  
\end{figure*}

The full stellar catalogue resulting from our analysis of the field
region contains 759 stars. Given that the cluster region covers 2.14
times that of the comparison field, from a statistical perspective we
therefore expect 1607 field stars to be located within the cluster
region. We divided the NGC 1651 cluster and field colour--magnitude
diagrams into 50 bins in magnitude and 25 bins in colour; for
relatively sparsely populated regions, we enlarged the bin size
appropriately (see below). We then calculated the number of field
stars in each colour--magnitude bin, and subsequently removed 2.14
times the (integer) number of stars from the corresponding bins of the
NGC 1651 colour--magnitude diagram.

Because the comparison field region was selected from the same image
as the cluster region, its exposure time is identical. Hence,
exposure-time differences will not affect the reliability of our
field-star decontamination, although statistical differences between
the cluster and field regions cause a slight dependence on the adopted
grid size. We carefully checked how the number of bins adopted would
affect the decontamination results and enlarged the bin sizes for
sparsely populated regions (e.g., on the red side of the main
sequence). We concluded that our field-star decontamination is robust
with respect to reasonable differences in adopted bin size. This thus
eventually resulted in a statistically robustly field-star
decontaminated colour--magnitude diagram of NGC 1651. We show the
results of the main steps employed in our field-star decontamination
procedure in Extended Data Fig. 2. Panel a shows the original
colour--magnitude diagram of NGC 1651 (for $R \le 75$ arcsec), panel b
represents the synthesized field-star equivalent and panel c is the
decontaminated colour--magnitude diagram on which we based our
analysis.

\section*{Using the subgiant branch to constrain the cluster's
    maximum age dispersion}

Many authors have invoked age dispersions as explanation for the
observed extended main-sequence turn-off regions, and although
numerous apparently somewhat different scenarios have been proposed,
most can be traced back to the basic idea of an age dispersion. For
instance, mergers of star clusters with an age difference of $\sim$200
Myr\cite{Mack07} as well as interactions of star clusters
and star-forming giant molecular clouds,\cite{Bekk09s} have been
suggested as the possible origin of extended turn-off regions.

We calculated the magnitude deviation ($\Delta B$) with respect to the
young, $\log[t \,({\rm yr})]  = 9.24$ isochrone for each
subgiant-branch star (cf. Fig. 2). Since our full sample contains 38
subgiant-branch stars, we adopted five bins in $\Delta B$. A gradually
increasing trend in $\Delta B$ is found, starting from $\Delta B \sim
-0.09$ mag, with a peak at $\Delta B \sim 0.00$ mag, followed by a
decrease to $\Delta B \sim 0.14$ mag and a slight upturn to $\Delta B
\sim 0.20$ mag: see Extended Data Fig. 3, which includes the 1$\sigma$
standard deviations. We next generated an additional set of isochrones
characterized by different ages and applied the same procedure. The
typical $\Delta B$ values are included at the top of Extended Data
Fig. 3 (black dashed lines), for an age resolution of $\Delta \log[t \,({\rm yr})] = 0.2$.

\begin{figure*}
  \centering
  \includegraphics[width=1.8\columnwidth]{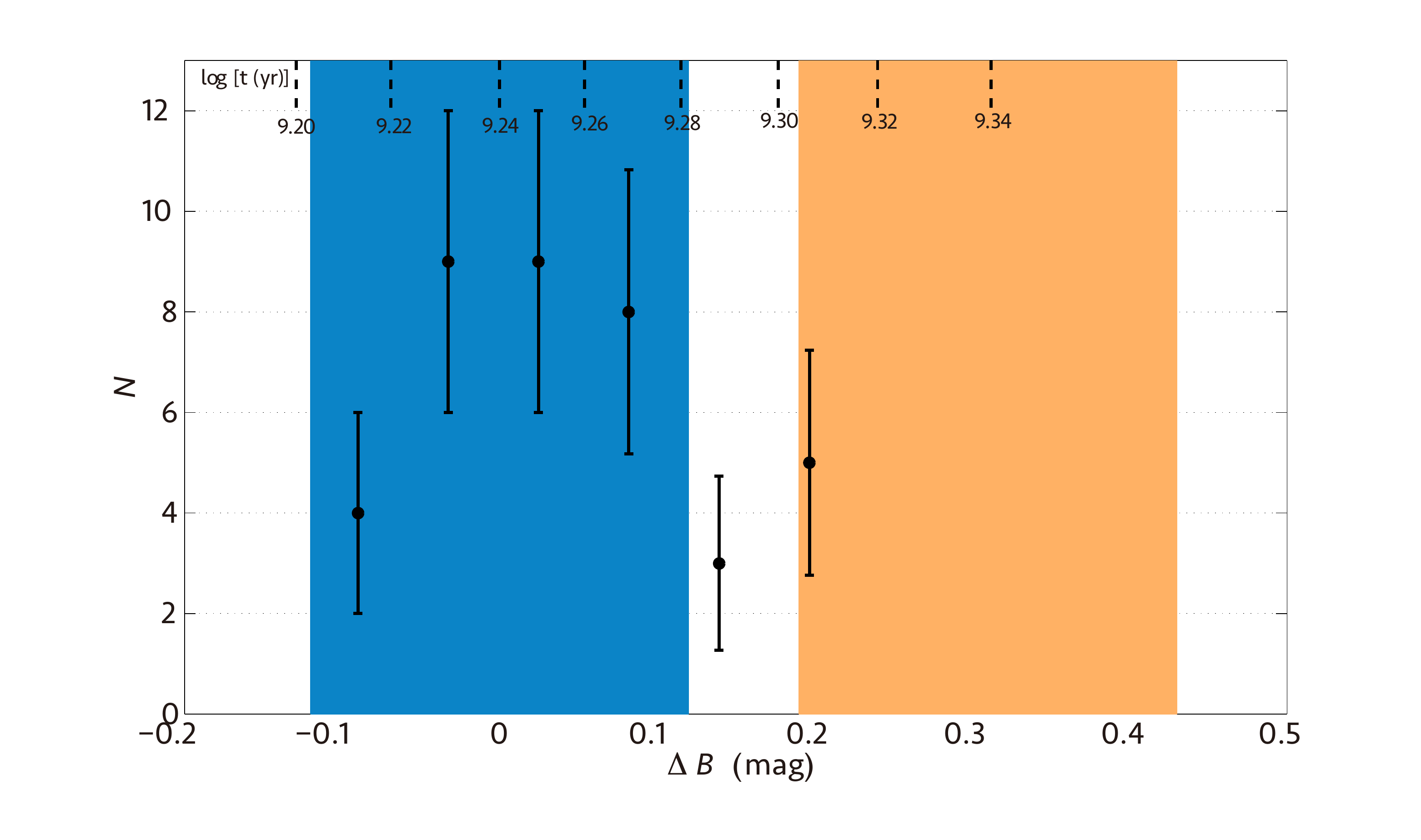}
\caption{{\bf Constraints on the maximum likely age dispersion.} Number distribution, $N$ (including 1$\sigma$ standard deviations), of the different ages, as indicated.
deviations in magnitude, $\Delta$B, of our subgiant-branch sample,  as in Fig. 2. The black dashed lines at the top indicate typical $\Delta B$ values for isochrones with differently ages, as indicated.}
    \label{S2}  
\end{figure*}

Assuming appropriate photometric uncertainties for each of these
isochrones, we calculated the number of subgiant-branch stars that
would be covered if we adopted a given age dispersion. We first
proceeded to test the simple-stellar-population approximation, i.e.,
assuming no age dispersion. In this case, all stars should be located
on the $\log[t \,({\rm yr})]  = 9.24$ isochrone, with a spread
determined by the typical (3$\sigma$) photometric uncertainties of
0.12 mag. We found that 30 of our 38 subgiant-branch stars are
associated with this isochrone (cf. Fig. 3, light blue background).

If we assume that all subgiant-branch stars belong to a simple stellar
population characterized by a typical age of $\log[t \,({\rm yr})] 
= 9.26$, and adopting the same photometric uncertainties, we can
reproduce 35 of the 38 stars (92\%). This thus strongly implies that
the NGC 1651 stellar population is most likely a genuine simple
stellar population. Extended Data Table 1 includes the results of our
analysis to derive the maximum intrinsic age dispersion needed to
explain the observed subgiant-branch loci in the cluster's
colour--magnitude diagram.

\begin{table}
\small
\caption{Age dispersions required to match the observed spread of subgiant-branch stars in NGC 1651}
\begin{center}
 \begin{minipage}{80mm}
    \vspace{2mm}
  \begin{tabular}{@{}cccc@{}}
  \hline
   $\Delta {\sf log}  [\textsf{\itshape t}\, ({\sf yr})]$ &  $\textsf{\itshape N}_{\sf SGB}$ & {\sf Fraction} (\%) & $\Delta \textsf{\itshape t}$ ({\sf Myr})\\
 \hline
    {\sf 9.24--9.28} & {\sf 38/38} & {\sf 100.0} & {\sf 167} \\
    {\sf 9.26--9.28} & {\sf 37/38} & {\sf 97.4} & {\sf 86} \\
    {\sf 9.24}--{\sf 9.26} & {\sf 36/38} & {\sf 94.7} & {\sf 82} \\
    {\sf 9.26} & {\sf 35/38} & {\sf 92.1} & {\sf SSP} \\
    {\sf 9.24} & {\sf 30/38} & {\sf 78.9} & {\sf SSP} \\
    {\sf 9.28} & {\sf 27/38} & {\sf 71.1} & {\sf SSP} \\
  \hline
    {\sf 9.24}--{\sf 9.26} & {\sf 15/15} & {\sf 100.0} &  {\sf 82} \\
    {\sf 9.26} & {\sf 14/15} & {\sf 93.3} &  {\sf SSP} \\
    {\sf 9.24} & {\sf 13/15} & {\sf 86.7} & {\sf SSP} \\
    {\sf 9.28} & {\sf 10/15} & {\sf 66.7} & {\sf SSP} \\
\hline
\end{tabular}
\end{minipage}
\end{center}
\end{table}

An age dispersion of $\sim$80 Myr can reproduce $>$90\% of
the subgiant-branch stars in our full sample. Similarly, if we assume
that the cluster's subgiant-branch stars are members of a simple
stellar population, a typical age of $\log[t \,({\rm yr})]  = 9.26$
can also reproduce $>$90\% of all subgiant-branch stars. The result
holds for the subgiant-branch sample in the cluster core: an age
dispersion of $\sim$80 Myr can reproduce all the core
subgiant-branch stars, and a simple-stellar-population model with a
typical age of $\log[t \,({\rm yr})]  = 9.26$ still reproduces
$>$90\% of the core subgiant-branch stars. This hence unequivocally
excludes the presence of an age dispersion extending to at least
$\log[t \,({\rm yr})]  = 9.34$.

\section*{A population of rapidly rotating stars?}

The observed extended main-sequence turn-offs in intermediate-age
clusters might also be explained as evidence of the presence of a
population of rapidly rotating stars.\cite{Bast09,Yang13,Li14}. The
centrifugal force resulting from rapid stellar rotation leads to a
reduction in effective gravity, which hence decreases both the stellar
surface temperature and its luminosity\cite{Yang13}. The reduced
gravity also leads to a decreasing stellar central hydrogen-burning
efficiency, rendering stars slightly fainter. This effect mainly
affects F-type stars; stars with masses below 1.2 solar masses do not
rotate rapidly because of magnetic braking\cite{Mest87s}.

Although some authors have claimed that rapid stellar rotation could
lead to a broadening of the turn-off\cite{Li12,Li14}, this scenario
only holds if rapid rotation does not have any effect on the stellar
lifetime on the main sequence. However, rapid rotation will also cause
a transfer of mass from radial shells to the central core, thus
providing additional material for nuclear fusion in the core. This
hence could increase the stellar lifetime compared with their
non-rotating counterparts. Calculations of the effect of this expected
prolongation of stellar lifetimes has led some authors\cite{Gira13} to
conclude that the resulting colour--magnitude diagram will still
retain a narrow main-sequence turn-off. These authors maintained that
the presence of an age dispersion was still the most natural model
that reproduces the extended turn-off. However, derivation of
colour--magnitude diagrams resulting from the adoption of different
rotation velocities,\cite{Yang13} while also considering the increased
main-sequence lifetimes, led to the conclusion that such a scenario
can still reproduce the observed extended turn-offs. However, the
extent of the main-sequence turn-off's broadening depends on the
typical cluster age. Nevertheless, if one adopts a modest mixing
efficiency for rotating stars, extended turn-offs can still be
observed.\cite{Yang13} In any case, since different stellar rotation
rates have been observed for solar-neighbourhood field
stars\cite{Roye13s}, it is natural to expect that stars in star
clusters may have similar distributions of rotation velocities.

Overall, the extent to which rapid rotation will affect
subgiant-branch stars is as yet unclear. Very few authors consider
these effects, with the exception of a single study\cite{Geor14} that
aims at generating a grid of stellar models including a range of
rotation rates. Although these authors have thus far only
satisfactorily completed their calculations for extremely massive
stars, using different evolutionary tracks and a range of rotation
velocities, this nevertheless allows us to estimate the extent to
which rapid rotation may affect stars on the subgiant branch. Based on
their interactive tools
(http://obswww.unige.ch/Recherche/evol/-Database-), we generated two
evolutionary tracks for their lowest-mass stars, 1.7 solar masses, one
without rotation and the second characterized by extremely rapid
rotation ($\omega$ = 0.95, i.e., rotation at 95\% of the critical
break-up rate): see Extended Data Fig. 4. We clearly see that,
following the main-sequence turn-off stage, the rapidly rotating track
converges to the non-rotating track. Indeed, because of the
conservation of angular momentum, the fast rotators are expected to
quickly slow down when they expand and evolve onto the subgiant
branch. This result hence confirms that the effects of rapid stellar
rotation become negligible, so that the observed narrow subgiant
branch in NGC 1651 can only be reconciled with the colour--magnitude
diagram of a genuine simple stellar population.

\begin{figure*}
  \centering \includegraphics[width=1.6\columnwidth]{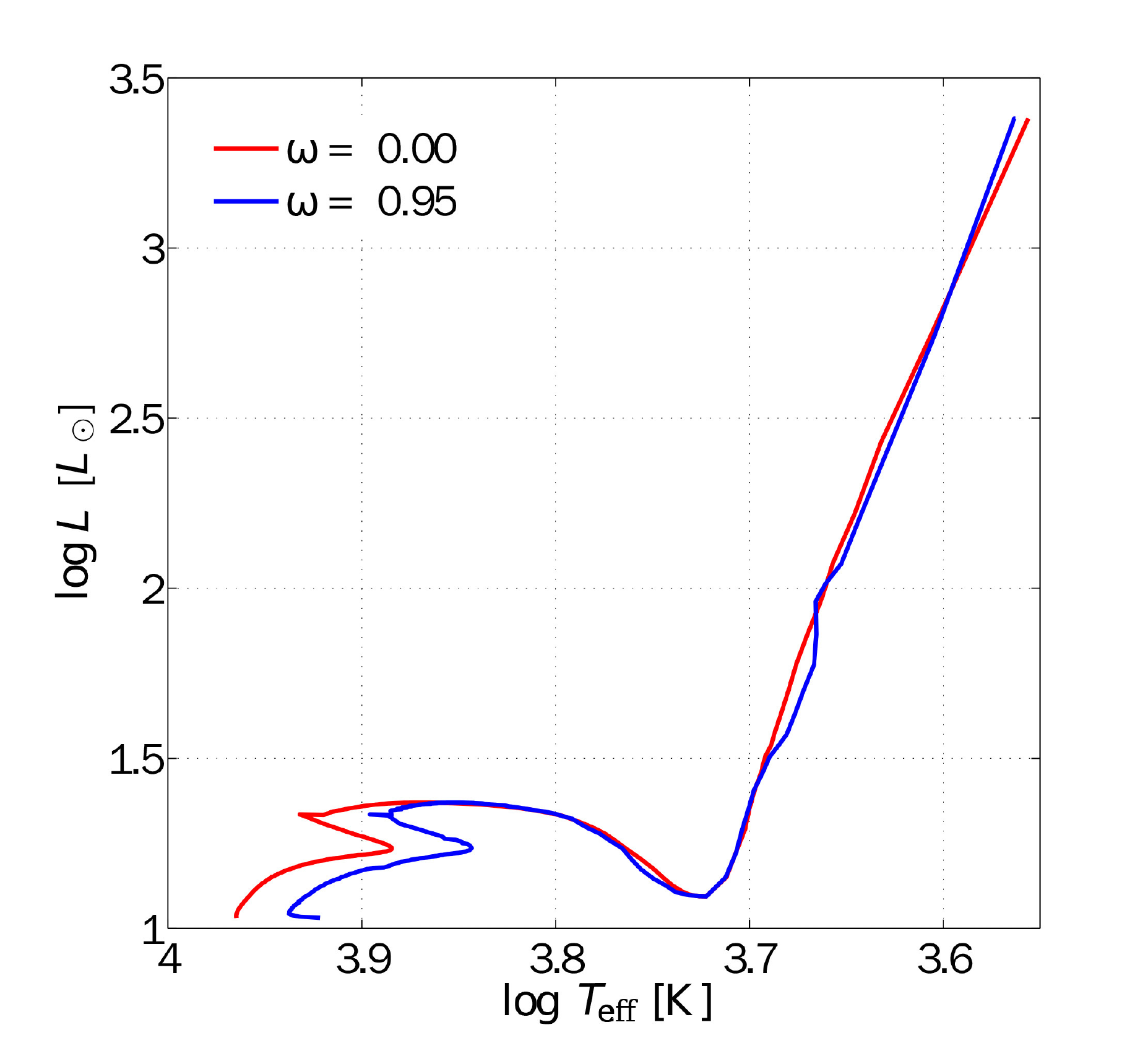}
\caption{{\bf Evolutionary tracks for extremes in stellar rotation
  rates.} Red: Non-rotating stars; blue: stellar rotation at 95\% of
  the critical break-up rate ($\omega = 0.95$). Both tracks apply to
  1.7 solar-mass stars. $L_\odot$: Solar luminosity. $T_{\rm eff}$:
  Effective temperature.}
    \label{S3}  
\end{figure*}

However, taking into account the effects of rapid rotation is highly
complex. Because stars that originally rotate rapidly tend to live
longer than their non-rotating counterparts, the presence of a
population of rapidly rotating stars may, in fact, still give rise to
a broadened or split subgiant branch. Whether or not this scenario
holds depends on the atmospheric mixing efficiency, the effects of
which are as yet unclear. Nevertheless, we point out that if the
mixing efficiency is reduced to ``normal'' levels of 0.03, the
extended main-sequence turn-off caused by the most rapid stellar
rotation will be equivalent to a simple-stellar-population age spread
of approximately 450 Myr for clusters aged 1.7 Gyr\cite{Yang13}. This
fits our observations exactly.

\section*{Additional evidence in support of NGC 1651 as a simple
  stellar population}

Except for possibly the cluster's sodium abundance, [Na/Fe], the
observed dispersions in the abundances of all other elements
investigated to date are consistent with the measurements'
root-mean-square values.\cite{Ales08s} [Na/Fe] ranges from
approximately $-0.41$ dex to $-0.03$ dex, but this result is based on
analysis of only five bright asymptotic giant-branch stars, which may
be strongly affected by their associated stellar winds. In fact, it
has been shown convincingly\cite{Ales08s,Mucc14s} that a number of
extended main-sequence turn-off clusters do not exhibit
chemical-abundance spreads.

Based on a detailed analysis of the spectra of 1200 red giants in 19
clusters\cite{Carr10s}, it has become apparent that first,
intermediate- and extreme second-generation stars tend to be found in
three typical zones in the [Na/Fe] versus [O/Fe] diagram. In this
context, first-generation stars may be characterized by relatively
poor sodium abundances, exhibiting dispersions of up to 0.4
dex\cite{Carr10s}. As such, the absence of any significant abundance
dispersions in most elements\cite{Ales08s} in the cluster, combined
with the observed spread in [Na/Fe], is indeed consistent with NGC
1651 representing a genuine simple stellar population.

Recent insights\cite{Bast14s} convincingly showed that star clusters
with ages up to 300 Myr in both Magellanic Clouds do not have any
sizeable gas reservoirs left to form second-generation stars. One must
thus turn to alternative models to explain the observations of
clusters like NGC 1651.

\end{document}